\documentclass[sigconf]{acmart}
\renewcommand\footnotetextcopyrightpermission[1]{} %

\usepackage{graphicx} %
\usepackage{orcidlink} %

\PassOptionsToPackage{hyphens}{url}\usepackage{hyperref}%

\usepackage[nameinlink, noabbrev, capitalize]{cleveref}%
\usepackage{soul}
\usepackage{xcolor}
\usepackage{subcaption}%
\usepackage[normalem]{ulem} %
\usepackage{enumitem}%

\usepackage{longtable}
\usepackage{array}

\usepackage{changepage}%
\renewenvironment{quote}
  {\begin{adjustwidth}{15pt}{0pt}\itshape}
  {\end{adjustwidth}}

\usepackage{xspace}
\newcommand{\ailab}{AI-Lab\xspace}
\newcommand{\ailabshort}{\ailab}
\title{A Scaffolded GenAI Lab in Early Undergraduate CS: A Mixed-Methods, Multi-Course Evaluation}
\date{March 2025}

\author{Ethan Dickey}
\authornote{Both authors contributed equally to this research.}
\affiliation{
  \institution{Purdue University}
  \streetaddress{Department of Computer Science}
  \city{West Lafayette}
  \state{Indiana}
  \country{USA}
  \postcode{47907}
  }
\email{dickeye@purdue.edu}
\orcid{0009-0007-3706-5253}
\author{Andres Bejarano}
\authornotemark[1]
\affiliation{
  \institution{Purdue University}
  \streetaddress{Department of Computer Science}
  \city{West Lafayette}
  \state{Indiana}
  \country{USA}
  \postcode{47907}
  }
\email{abejara@purdue.edu}
\orcid{0000-0003-2611-2855}
\author{Rhianna Kuperus}
\affiliation{
  \streetaddress{Office of Institutional Research, Assessment and Effectiveness}
  \institution{Purdue University}
  \city{West Lafayette}
  \state{Indiana}
  \country{USA}
  \postcode{47907}
  }
\email{rsetsma@purdue.edu}
\orcid{0009-0002-3057-7671}
\author{Bárbara Fagundes}
\affiliation{
  \streetaddress{Center for Instructional Excellence}
  \institution{Purdue University}
  \city{West Lafayette}
  \state{Indiana}
  \country{USA}
  \postcode{47907}
  }
\email{bfagundes@purdue.edu}
\orcid{0000-0001-8138-1471} %
\begin{abstract}
    \textbf{Background and Context.} Generative AI (GenAI) tools are increasingly used in programming courses, but we have limited evidence about how brief instruction can foster responsible, learning-oriented use.
    
    \textbf{Objectives.} We evaluate \ailab%
    , a scaffolded GenAI literacy intervention, asking how students' self-reported GenAI usage and their openness and comfort using GenAI for conceptual, debugging, and homework tasks change after participation.
    
    \textbf{Method.} Across two semesters in three CS courses and one first-year engineering course at a U.S. university, we deployed the \ailab (pre-lab orientation, in-class critique of GenAI outputs, and a required homework reflection), collecting paired pre/post surveys (Perception N=831; Usage N=826) and six post-intervention focus groups; primary inferential analyses used the three CS courses (N=778 and 773, respectively). We analyzed survey shifts with paired non-parametric tests and focus groups via thematic analysis.
    
    \textbf{Findings.} Openness increased for conceptual questions and homework help, and comfort increased for conceptual, debugging, and homework scenarios; self-reported frequency of GenAI use for homework and projects remained stable, while self-reported use for debugging increased. Focus group participants described adopting more iterative prompting strategies, becoming more skeptical of correctness, and articulating clearer boundaries around integrity and dependence.
    
    \textbf{Implications.} A short, structured intervention can shift students' reported comfort with and willingness to use GenAI and influence the strategies they describe for engaging with it without increasing overall self-reported use on graded work. These results motivate future work triangulating surveys with behavioral traces and learning measures.
    
\end{abstract}

\ccsdesc[500]{Social and professional topics~Computing education}

\keywords{Generative AI (GenAI), core skill development, pedagogical framework, \ailab, mixed methods (quantitative and qualitative)}

\begin{document}

\maketitle
\pagestyle{plain}%
\section{Introduction}
The widespread availability of generative AI (GenAI) tools has disrupted computing coursework by enabling students to generate code and explanations on demand, raising new pedagogical and ethical challenges \cite{tlili2024can, donaire2025tiza, lau2023BanIt, Becker2023programmingIsHard, Finnie22, Moradi23}. As these tools become normalized, educators face pressure to clarify what learning goals remain central and what instructional supports students need to use GenAI in ways aligned with learning and integrity expectations \cite{Kerslake10114536415543701934, Hassan10114536415543701906, Raihan2025}.
Existing guidance emphasizes helping students evaluate GenAI outputs and discuss ethical and integrity implications, but many instructors lack the bandwidth for course-wide redesign \cite{Bull23}.
Several studies also highlight the necessity of educating students on the ethical implications and limitations of GenAI \cite{ozkaya_2024, Generative_Ai_2025_scholarlyteacher}. Key concerns include safeguarding sensitive information, recognizing the potential inaccuracies of AI-generated content, understanding AI's capacity to produce harmful material, and adhering to platform-specific terms of use.

In this study, we focus on a pragmatic instructional problem: many instructors want to acknowledge that GenAI exists and is widely used, but do not have room to redesign an entire course around GenAI or build extensive new assessment infrastructure. We therefore evaluate a brief, course-embedded scaffold -- the \ailab
\cite{dickey2024innovating} -- that can be implemented within an existing course schedule. Rather than asking whether GenAI ``helps'' or ``harms'' learning in general, we examine what shifts are visible in students' self-reported usage patterns and their openness and comfort using GenAI across common academic scenarios, complemented by focus group accounts of how students describe using and bounding these tools.

\textbf{Contributions.} This paper makes three contributions to computing education research and practice:
\begin{enumerate}
    \item a mixed-methods, multi-course evaluation of a brief scaffolded GenAI lab intervention using paired pre/post surveys and focus groups;
    \item evidence of where students' self-reported perceptions and practices shifted and where they remained stable, including null findings that are pedagogically consequential.
    \item an implementable intervention structure that instructors can adapt, including timing, components, and example activities, with explicit boundary conditions and threats to validity (notably reliance on self-report and no learning outcome measures).
\end{enumerate}

\subsection{Related Work}
Recent advances in GenAI -- and especially large language models that can generate and explain code -- are reshaping the ``help'' landscape available to computing students and complicating long-standing assumptions about what it means to learn programming and how that learning should be assessed. Broad community-facing work has outlined both the urgency of this shift and the range of possible instructional responses, including changes to assessment design, increased attention to evaluation and verification skills, and explicit discussion of responsible use \cite{denny2024computing}.
At the same time, the CS education research literature on LLMs is growing so quickly that recent syntheses have begun cataloging emerging teaching practices, tools, and research findings across venues. In particular, systematic reviews and working group reports summarize an increasingly diverse set of interventions -- ranging from course redesigns to tool-mediated feedback -- while also highlighting persistent uncertainty about instructional goals and evaluation approaches \cite{Raihan2025}.

A large portion of early work focuses on course- and curriculum-level adaptation, especially in introductory programming. One response has been to publish ``emerging practice'' guidelines for how CS1 instructors might adapt assignments, assessment practices, and classroom norms as GenAI becomes more common \cite{mahon2024guidelines}. Another response has been more structural: redesigning course learning objectives to explicitly incorporate LLM-mediated programming workflows. For example, \cite{Vadaparty2024} describes a CS1 course that embraces LLMs from the beginning and intentionally shifts emphasis away from writing code from scratch toward decomposition, explanation, testing, and evaluation -- skills positioned as necessary to produce correct software when an LLM is part of the workflow. These examples illustrate a common theme in the literature: the presence of GenAI pressures educators to reconsider whether ``syntax production'' remains a primary bottleneck, and instead to foreground higher-level reasoning and evaluation activities \cite{denny2024computing}. 

A complementary strand examines what students and instructors are already doing with GenAI, and what risks and inequities may follow from those practices. Interviews with university programming instructors show movement from early ``ban'' impulses toward more varied plans -- including restrictions, conditional permissions, and redesign, often driven by concerns about academic integrity and uncertainty about which learning outcomes can still be assessed with conventional tasks \cite{lau2023BanIt}. On the student side, early snapshots suggest substantial variation in how and why students adopt GenAI and what support they want from instructors. For instance, an ITiCSE study of early adoption captured student-reported use cases and concerns in a period before many institutional policies had stabilized \cite{smith2024early}. Work on trust likewise suggests that students' reliance on GenAI is shaped by beliefs about its reliability and appropriateness for coursework, reinforcing that ``tool availability'' does not translate into uniform or pedagogically beneficial use \cite{amoozadeh2024trust}. Finally, empirical work suggests that GenAI may not reduce struggle equally for all novices; rather, it can amplify differences in students' ability to evaluate and correct generated artifacts. The ``Widening Gap'' study, for example, documents both benefits and harms for novice programmers and explicitly motivates the need for scaffolding that supports more effective and metacognitively aware use \cite{prather2024widening}. In a related direction, journal evidence indicates that students without sufficient background can have difficulty judging the quality of AI-generated answers, underscoring the importance of explicit instruction in evaluation rather than treating GenAI access as inherently supportive \cite{shoufan2023can}.

A third strand investigates explicit instruction aimed at helping students use GenAI in more learning-oriented ways, particularly beyond CS1. For example, ``Unlocking Potential with Generative AI Instruction'' studied the impact of incorporating instruction on industry-standard GenAI tools into a mid-level software development course and examined changes in students' perceptions, behaviors, and adoption \cite{Benario2025}. Alongside such instructional interventions, other work explores GenAI as part of the instructional infrastructure itself: LLMs have been used to generate practice problems and contextualized exercises, aiming to reduce instructor workload and expand available practice opportunities \cite{Logacheva10114536326203671103}. LLMs have also been explored as sources of automated formative feedback, though empirical findings caution that unmonitored feedback can include false positives/negatives and may require students to adopt verification behaviors (e.g., using test feedback) to benefit reliably \cite{Kerslake10114536415543701934}.%

Despite this growing body of work, less is known about lightweight, course-agnostic interventions that can be embedded within existing computing courses (including early- and mid-level courses) to scaffold responsible and learning-oriented GenAI use without requiring a full course redesign. In this paper, we examine the \ailab as a scaffolded, transferable lab activity intended to help students develop practical strategies for using GenAI in ways aligned with course learning goals and academic integrity expectations, and we evaluate its relationship to changes in students' reported use patterns and stances toward GenAI across common learning scenarios.

\subsection{AI Literacy}
As GenAI tools become embedded in students' everyday learning workflows, many educators have adopted the language of AI literacy to describe the capabilities students need to participate safely and effectively in AI-mediated environments. A widely cited definition frames AI literacy as a set of competencies that enable people to critically evaluate AI technologies, communicate and collaborate effectively with AI, and use AI as a tool in everyday contexts \cite{Long2020}. However, literature reviews emphasize that ``AI literacy'' has been used to describe a wide range of learning goals, instructional contexts, and levels -- from technical understanding of AI mechanisms to practical tool-use skills and to sociocultural critique -- making the construct difficult to operationalize consistently \cite{NG2021}. Recent CS education work further argues that, in the wake of GenAI, the term has become even more expansive and ambiguous, and that the field needs more specialized language to describe the specific literacies targeted by particular interventions \cite{Gu2025}.
One useful way to refine the construct is to distinguish (a) the objective of the literacy and (b) the perspective on AI that the literacy emphasizes. An integrative review of AI literacy and GenAI work identifies multiple literacy objectives -- such as functional tool use versus more critical evaluative stances -- as well as perspectives that range from technical detail to tool-oriented interaction to sociocultural framing \cite{Gu2025}. In parallel, practitioner-facing frameworks (e.g., in K-12 settings) similarly break AI learning goals into interpretable dimensions (such as social/ethical impacts and practical applications), illustrating that ``AI literacy'' is often best understood as an umbrella term whose meaningful use depends on contextual narrowing \cite{sentance2022perspectives}.

In programming courses, the rapid uptake of LLM-based assistants motivates a programming-situated GenAI literacy focused less on model internals and more on practical competencies that support learning-aligned use. Experience-report and integration work highlights recurring competencies such as decomposing problems into solvable subproblems, articulating intent and constraints clearly, and validating generated code and explanations through testing and inspection \cite{Vadaparty2024}. At the same time, evidence on novice interactions suggests that tool access alone does not guarantee productive use: students may over-trust plausible outputs or repeatedly request ``fixes'' without developing understanding, patterns associated with reduced metacognitive engagement and widening performance gaps \cite{prather2024widening}. Accordingly, GenAI literacy for programming also includes managing uncertainty and making principled decisions about when \emph{not} to use GenAI, aligning tool use with course norms and academic integrity expectations \cite{lau2023BanIt}.

The \ailab is designed as a scaffolded, low-overhead intervention that targets this programming-situated subset of AI literacy: practical strategies for interacting with GenAI, critically evaluating outputs, and making responsible choices about when and how to use GenAI in common learning tasks. This focus aligns with recent calls to operationalize ``AI literacy'' more narrowly and transparently in order to evaluate interventions meaningfully \cite{Gu2025}. Accordingly, rather than claiming to measure AI literacy as a global construct, we examine how participation in the \ailab relates to changes in students' reported behaviors (how they say they use GenAI in typical tasks) and their reported stances (openness and comfort using GenAI in those tasks).

\subsection{Research Questions}\label{sec:intro:RQs}
Motivated by the rapid normalization of GenAI use in computing coursework and by evidence that productive use depends on evaluation and boundary-setting skills that are not guaranteed by tool access alone, we investigate the \ailab as a scalable, course-agnostic approach to scaffolding learning-oriented GenAI use.
Building on prior work that introduced \ailab as a practical instructional framework and reported early course deployments \cite{dickey2024innovating,Bejarano10114536415553705201}, we examine what shifts are observable when the intervention is deployed across multiple early undergraduate course contexts. Prior work has documented instructor uncertainty and diverse policy responses, course redesigns that embed LLM workflows, and targeted instruction within specific courses; however, there remains limited evidence about lightweight interventions that can be incorporated across course contexts and still shape students' day-to-day GenAI practices and scenario-specific attitudes. 
We therefore ask:
\begin{enumerate}
    \item[(RQ1)] How and how much are students using GenAI before and after the \ailab intervention?
    \item[(RQ2)] Are students open to and comfortable with using GenAI in their learning processes? If so, in what scenarios?
\end{enumerate}

\section{Methods}
We used a mixed-methods design to examine how students reported using GenAI and how their reported comfort and openness to GenAI shifted following a brief scaffolded instructional intervention (the \ailab). Quantitative data consisted of paired pre/post surveys administered immediately around the intervention. Qualitative focus groups were conducted post-intervention to contextualize and interpret reported survey shifts (qualitative procedures are described in the subsequent subsections).

\subsection{Study Context and Courses}\label{sec:results:student_populations}
The study was conducted at a large U.S. research university across two semesters (Spring 2024 and Fall 2024) in three early undergraduate computer science courses and one first-year engineering course. Not all courses ran in both semesters. We use the following course labels throughout: DSA-CS (a CS2-level Data Structures and Algorithms course for CS majors; offered in both semesters), DSA-DSAI (a parallel CS2-level Data Structures and Algorithms course for Data Science/AI majors with some topic emphasis differences; offered in Spring), CP (an elective Competitive Programming I course typically taken after CS2; offered in Fall), and ENGR (a first-year engineering course emphasizing design and systems thinking; offered in Fall). Unless noted otherwise, our primary analyses focus on the computer science courses (DSA-CS, DSA-DSAI, and CP), with the engineering course reported separately where included.

\subsection{\ailab}
The \ailab is a scaffolded instructional sequence that makes GenAI's benefits and limitations visible in course-relevant contexts \cite{dickey2024innovating}. The intervention consists of: (1) a brief pre-lab orientation covering course expectations, integrity boundaries, and basic prompting; (2) an in-class activity in which students observe and critique GenAI outputs on course-aligned topics designed to yield plausible but incomplete or incorrect responses; and (3) a post-lab assignment add-on requiring students to document and reflect on their interaction with GenAI on a specified homework (e.g., how prompts were refined, what was verified, and where the model failed). Instructors can adapt prompts and post-lab tasks to local course content while maintaining the scaffolded structure.
\Cref{tab:ailab_summary} summarizes how the \ailab was implemented in each offering, including timing within the term, session duration, topical focus, deliverables, and workload features.
\begin{table*}%
    \resizebox{\linewidth}{!}{
        \begin{tabular}{l|l|l|l|l}
            Course                                    & CP   & DSA-CS  & DSA-DSAI & ENGR \\ \hline
            
            Semester                                  & F24  & S24/F24 & S24      & F24 \\
            Week of term                              & 9    & 10/10   & 11       & 14 \\
            Content                                   & SCCs & SCCs    & UF/DS    & $\star$ \\
            In-class lab duration (min)               & 75m  & 50m     & 50m      & 80m \\
            \shortstack[l]{Postlab (assignment)\\~\\~}   &
                    \shortstack[l]{Use unlimited on $\ge$1\\problem + reflect\\~} &
                    \shortstack[l]{Only use GenAI\\until failure,\\report + reflect} &
                    \shortstack[l]{Only use GenAI\\until failure,\\report + reflect} &
                    \shortstack[l]{Post-lecture\\reflection\\~} \\

            \shortstack[l]{Grading (participation/\\
            reflection vs correctness)}               & Refl. + Correctness & Refl. + Completion & Refl. + Completion & Refl. only \\\cline{2-5}
            \shortstack[l]{Workload footprint\\(instructors/TAs)} & \multicolumn{4}{l}{\shortstack[l]{Prelab/survey prep (30m); lecture prep (30m-1hr);\\grading (avg 30s-1m/student)}}
        \end{tabular}
    }
    \caption{\ailab implementation summary by course offering (timing, duration, components, and student deliverables). All courses suggested using ChatGPT but students were free to choose any that had a ``share'' feature. An example set (prelab/in-class outline/postlab) and other instructional materials can be found in Appendix \ref{app:focus_group_questionnaire_and_lab_artifacts}. See \Cref{sec:results:student_populations} for course acronyms and context. SCCs = Strongly Connected Components; UF/DS = Union Find/Disjoint Sets; $\star$ = Individual creative programming project brainstorming.}
    \label{tab:ailab_summary}
\end{table*}

\subsection{Survey Design and Data Collection}
We administered paired pre/post surveys immediately before and after the \ailab intervention to measure changes in students' self-reported GenAI practices and perceptions. The survey battery consisted of two instruments administered at both timepoints: a Usage survey (reported frequency of GenAI use across common academic scenarios) and a Perception survey (reported openness and comfort using GenAI across scenarios). Items were scenario-based and aligned to tasks students commonly encounter (e.g., conceptual questions, debugging, and homework problems). Response scales were ordinal (Likert-type).

\subsubsection{Participants and inclusion criteria.}
All students enrolled in participating course offerings were invited to complete the surveys. For paired pre/post comparisons, we included only responses from students who completed both the pre- and post-intervention survey for the relevant instrument; therefore, pre and post counts are identical by construction within the paired sample. 
See \Cref{tab:combined_perception_usage_counts} for paired sample sizes by course and instrument.

\begin{table}
    \begin{tabular}{lll}
        \textbf{Course}      & \textbf{Paired N (Perception)} & \textbf{Paired N (Usage)} \\
        CP                   & 43                             & 43                        \\
        DSA-CS               & 676                            & 671                       \\
        DSA-DSAI             & 59                             & 59                        \\\hline
        \textbf{Total}       & \textbf{778}                   & \textbf{773}              \\%
        ENGR                 & 53                             & 53                        \\\hline
        \textbf{Grand Total} & \textbf{831}                   & \textbf{826}      
    \end{tabular}
    \caption{Perception and Usage survey totals, combined across courses and semesters. See \Cref{tab:combined_perception_semester_counts,tab:combined_usage_semester_counts} for a detailed breakdown across semesters and courses.}
    \label{tab:combined_perception_usage_counts}
\end{table}

\subsection{Statistical Analysis}
\subsubsection{Overview and outcome measures}
The quantitative component uses paired pre/post survey responses to examine changes in students' self-reported GenAI usage and perceptions following the \ailab. The Perception survey includes scenario-based items measuring (a) openness to using GenAI and (b) comfort using GenAI in common course contexts (conceptual questions, debugging, homework). The Usage survey includes items measuring the frequency with which students report using GenAI in comparable contexts (conceptual questions, debugging, homework, programming projects). Response options are ordinal Likert-type scales (5-point for openness/comfort; 4-point for frequency). For each item, we compare students' post-intervention response to their pre-intervention response, yielding paired observations.

\subsubsection{Paired pre/post comparisons}
Because the survey responses are ordinal and paired within individuals, we use Wilcoxon signed-rank tests to assess whether the distribution of paired responses differs between pre and post for each item. Tests are two-tailed and are conducted separately for each survey item. Consistent with standard Wilcoxon procedures, pairs with zero difference (no pre/post change) do not contribute to the signed-rank statistic for that item. For each paired survey item, we tested $H_0: \tilde{d} = 0$ versus $H_a: \tilde{d}\ne 0$, where $\tilde{d}$ is the median of the paired differences.

\subsubsection{Effect size estimation with rank-biserial correlation}
In addition to statistical significance, we report an effect size for paired ordinal comparisons using rank-biserial correlation ($r_{rb}$). Rank-biserial correlation summarizes the extent to which post-intervention responses tend to be higher or lower than pre-intervention responses in the paired sample.
For each item, we derive $r_{rb}$ from the Wilcoxon signed-rank components (i.e., the balance of ranked positive vs. negative paired differences). As with the signed-rank test, paired observations with no change (ties) do not contribute to the signed-rank computation; accordingly, $r_{rb}$ reflects the direction and magnitude of change among respondents who shifted their rating.

Positive values of $r_{rb}$ indicate a tendency toward higher post-intervention responses, negative values indicate a tendency toward lower post-intervention responses, and values near zero indicate little directional shift.

To support interpretability, we use the following descriptive magnitude bands for $r_{rb}$: negligible (|0.00-0.10|), small (|0.10-0.30|), medium (|0.30-0.50|), and large (>|0.50|). Because magnitude labels can be sensitive to context and sample size, we treat these categories as descriptive rather than as claims of educational importance on their own.

\subsubsection{Significance threshold and reporting}
We use a significance threshold of $\alpha=0.05$. For each item, we report the Wilcoxon signed-rank statistic and p-value along with the corresponding $r_{rb}$. We also report descriptive pre/post response distributions (counts and percentages) to enable readers to interpret the direction and practical meaning of shifts. Because multiple items are tested, we emphasize the overall pattern of results and interpret individual p-values cautiously rather than treating any single item as definitive. All itemwise p-values are unadjusted and exploratory.

\subsection{Focus Groups and Data Collection}\label{sec:methods:qual}
To contextualize survey shifts, we conducted six semi-structured focus groups after the \ailab intervention. Participants were recruited via email from participating course offerings; participation was voluntary and compensated. Sessions were held approximately 5-7 weeks after the intervention to allow students time to apply GenAI in subsequent coursework; this supports reflection beyond immediate post-lab impressions but may reduce recall of specific lab details. Focus groups followed a common protocol (Appendix \ref{app:focus_group_questionnaire_and_lab_artifacts}), were audio-recorded with consent, transcribed, and de-identified prior to analysis. A researcher facilitated discussion to elicit students' accounts of how they used GenAI before/after the \ailab, how they evaluated correctness, and how they reasoned about boundaries and integrity.

\subsection{Qualitative Analysis}\label{sec:methods:qual_analysis}
To explore how students engaged with GenAI tools before and after the \ailab intervention, we conducted a thematic analysis of post-intervention focus group transcripts \cite{braun2006thematic, saldana2013manual}. Our goal was to capture changes in perception, behavior, and strategic thinking about GenAI usage that may not have been as clear through the surveys.

We began coding deductively, drawing categories from our research questions and the focus group protocol. Two researchers independently coded two transcripts using this initial framework and then met to compare interpretations and resolve discrepancies. This dialogue helped us refine our code definitions and build alignment for the next stages. We proceeded to code the remaining transcripts in NVivo, applying the refined codebook while looking for new patterns that emerged. Throughout, we kept analytic memos to capture reflections and guide the development of themes.

After coding, we grouped related codes into broader themes that captured the patterns and perspectives present in the transcripts. Our intention was not to generalize to all students, but to surface insights grounded in the data. The themes reflect both shared experiences and nuanced tensions in how students made sense of GenAI in their coursework.

To support trustworthiness, we involved multiple coders, revisited our interpretations over multiple rounds, and used quotes to illustrate each theme. \Cref{tab:qualthemes} summarizes the final themes, the codes that informed them, and example quotes from students in different focus groups. These themes are further discussed in \Cref{sec:results:qual_analysis}, and the focus group protocol is included in Appendix \ref{app:focus_group_questionnaire_and_lab_artifacts}.

\begin{table*}
    \caption{Representative quotes supporting each qualitative theme.}
    \label{tab:qualthemes}
    \begin{tabular}{|>{\raggedright\arraybackslash}p{1.75cm}|>{\raggedright\arraybackslash}p{2.25cm}|p{12.25cm}|}
        \hline
        \textbf{Theme} & \textbf{Related Codes} & \textbf{Representative Quotes (Focus Group X, Speaker Y)} \\ \hline
        
        From Casual to Strategic Use & Post-intervention use, Prompting strategies &
        ``Now I give it, for example, everything I do know... Based on what I have, give me the answer to A.'' - FG6, S2 \par
        ``Instead of just asking one question and moving on, I now try to refine my questions to make sure I'm getting the best answer.'' - FG1, S2 \par
        ``I used to just throw a question at it, and now I engage with it in a more back-and-forth way. I give it more context.'' - FG5, S3 \\
        \hline
        
        Stable Frequency, Evolved Rationale & Pre-/Post- frequency, Use consistency &
        ``It's pretty much the same. I just learned better ways to use it.'' - FG1, S3 \par
        ``The frequency stayed the same, but now I give it everything I do know... I tell ChatGPT, don't change those answers.'' - FG6, S2 \par
        ``I don't know if I've used it more, but I definitely use it differently.'' - FG5, S3 \\
        \hline
        
        Improved Awareness of AI Limitations & Accuracy concerns, Limitations of GenAI &
        ``It would make it look really, really right... but the definition itself is completely wrong.'' - FG5, S4 \par
        ``In my experience, it's not super great at math... probability, linear algebra... not too great at.'' - FG1, S2 \par
        ``You just told me... I cannot access articles through links. It was just funny how... really obvious things beyond its capability.'' - FG6, S2 \\
        \hline
        
        GenAI as a Support Tool & Scenarios students are willing to use AI &
        ``It's sort of like questions I'd ask a TA. I just ask Chat first.'' - FG3, S5 \par
        ``I basically use it to break down hard concepts into simpler terms.'' - FG6, S3 \par
        ``I primarily use it for research... it's been really good for taking notes.'' - FG5, S2 \\
        \hline
        
        Avoidance and Boundaries & Scenarios not to use AI, Ethical hesitation &
        ``Probably on projects or homework... there's a risk of an AI detection being run.'' - FG1, S3 \par
        ``If your code is very complex, sometimes it just doesn't know where to start.'' - FG5, S2 \\
        \hline
        Ethical Concerns and Internal Tension & Internal tension, Integrity concerns &
        ``I reduced using it... I just feel like I'm getting more dumb... a puppet of ChatGPT.'' - FG5, S4 \par
        ``The biggest thing is probably... ethics... I don't want to get an advantage over my peers.'' - FG1, S2 \par
        ``I don't want it to just generate content for me. I want to make sure I'm still learning.'' - FG1, S3 \\
        \hline
    \end{tabular}
\end{table*}

\section{Results}
We first individually cover quantitative and qualitative results, then present additional descriptive results that were not able to be included in the statistical analysis.

\subsection{Statistical Analysis Results}\label{sec:results:quant_analysis}
\paragraph{Summary of pre/post survey shifts.} Across the paired survey sample, students reported increased comfort using GenAI for conceptual questions, debugging, and homework scenarios following the \ailab. Students also reported increased openness to using GenAI for conceptual questions and for homework help. In contrast, self-reported frequency of GenAI use for homework and programming projects did not change, while self-reported use for debugging increased. We interpret these results as evidence that the intervention is associated with shifts in students' reported comfort and willingness, with a more targeted change in reported debugging use rather than a broad increase in reported use on graded work.

Tables \ref{tab:combined:open:conceptual}--\ref{tab:combined:use:projects} show the results of surveys in S24 and F24 across DSA-CS, DSA-DSAI, and CP, followed by their statistical analysis. See \Cref{tab:combined_perception_usage_counts}
for distribution of students across courses.
\Cref{tab:pvalues_combined_no_ENGR} summarizes the two-tailed Wilcoxon signed-rank results and effect sizes for each item; we treat items with $p<0.05$ as evidence of a pre-post shift. Six of nine items showed statistically detectable pre/post shifts in the paired CS sample, indicating strong evidence to reject the null hypothesis (there is no difference) for most of the questions.

\paragraph{Exploratory subgroup checks.} We examined whether pre/post shifts differed by student demographics (gender, race/ethnicity, residency status, and class standing). We did not detect statistically significant differences in the direction or magnitude of shifts across these groups. Because these checks were exploratory and some subgroups were comparatively small, we interpret this result as no differences detected rather than evidence of equivalence.

\begin{table*}%
    \caption*{Openness Tables \ref{tab:combined:open:conceptual}--\ref{tab:combined:open:homework}. Scale: 1-5: Very Opposed, Somewhat, Indifferent, Somewhat, Very Open. N=778.}
    \parbox{0.32\linewidth}{
        \caption{How open are you to using GenAI to get help with conceptual questions? \label{tab:combined:open:conceptual}}
        \begin{tabular}{lllll}
              & Before       & \%               & After        & \%               \\
        1     & 20           & 2.6\%            & 13           & 1.7\%            \\
        2     & 57           & 7.3\%            & 39           & 5.0\%            \\
        3     & 66           & 8.5\%            & 57           & 7.3\%            \\
        4     & 241          & 31.0\%           & 255          & 32.8\%           \\
        5     & 394          & 50.6\%           & 414          & 53.2\%           
        \end{tabular}
    }
    \hfill
    \parbox{0.32\linewidth}{
        \caption{How open are you to using GenAI to get help with debugging? \label{tab:combined:open:debugging}}
        \begin{tabular}{lllll}
              & Before       & \%               & After        & \%               \\
        1     & 35           & 4.5\%            & 30           & 3.9\%            \\
        2     & 72           & 9.3\%            & 78           & 10.0\%           \\
        3     & 140          & 18.0\%           & 114          & 14.7\%           \\
        4     & 275          & 35.3\%           & 316          & 40.6\%           \\
        5     & 256          & 32.9\%           & 240          & 30.8\%           
        \end{tabular}
    }
    \hfill
    \parbox{0.32\linewidth}{
        \caption{How open are you to using GenAI to get help with homework problems? \label{tab:combined:open:homework}}
        \begin{tabular}{lllll}
              & Before       & \%               & After        & \%               \\
        1     & 71           & 9.1\%            & 64           & 8.2\%            \\
        2     & 210          & 27.0\%           & 180          & 23.1\%           \\
        3     & 191          & 24.6\%           & 202          & 26.0\%           \\
        4     & 201          & 25.8\%           & 230          & 29.6\%           \\
        5     & 105          & 13.5\%           & 102          & 13.1\%           
        \end{tabular}
    }
\end{table*}

\begin{table*}%
    \caption*{Comfort in Use Tables \ref{tab:combined:comfort:conceptual}--\ref{tab:combined:comfort:homework} scale: 1-5: Very Uncomfortable, Somewhat, Indifferent, Somewhat, Very Comfortable. N=778.}
    \parbox{0.32\linewidth}{
        \caption{How comfortable are you using GenAI to help with conceptual questions? \label{tab:combined:comfort:conceptual}}
        \begin{tabular}{lllll}
              & Before       & \%               & After        & \%               \\
        1     & 28           & 3.6\%            & 21           & 2.7\%            \\
        2     & 64           & 8.2\%            & 41           & 5.3\%            \\
        3     & 121          & 15.6\%           & 91           & 11.7\%           \\
        4     & 290          & 37.3\%           & 329          & 42.3\%           \\
        5     & 275          & 35.3\%           & 296          & 38.0\%           
        \end{tabular}
    }
    \hfill
    \parbox{0.32\linewidth}{
        \caption{How comfortable are you using GenAI to help with debugging? \label{tab:combined:comfort:debugging}}
        \begin{tabular}{lllll}
              & Before       & \%               & After        & \%               \\
        1     & 66           & 8.5\%            & 42           & 5.4\%            \\
        2     & 119          & 15.3\%           & 123          & 15.8\%           \\
        3     & 200          & 25.7\%           & 166          & 21.3\%           \\
        4     & 224          & 28.8\%           & 284          & 36.5\%           \\
        5     & 169          & 21.7\%           & 163          & 21.0\%           
        \end{tabular}
    }
    \hfill
    \parbox{0.32\linewidth}{
        \caption{How comfortable are you using GenAI to help with homework problems? \label{tab:combined:comfort:homework}}
        \begin{tabular}{lllll}
              & Before       & \%               & After        & \%               \\
        1     & 93           & 12.0\%           & 87           & 11.2\%           \\
        2     & 196          & 25.2\%           & 176          & 22.6\%           \\
        3     & 228          & 29.3\%           & 200          & 25.7\%           \\
        4     & 157          & 20.2\%           & 225          & 28.9\%           \\
        5     & 104          & 13.4\%           & 90           & 11.6\%           
        \end{tabular}
    }
\end{table*}

\begin{table*}%
    \caption*{Frequency of Use Tables \ref{tab:combined:use:conceptual}--\ref{tab:combined:use:projects} scale: 1-4: Never, On Occasion, Moderately Frequently, Very Frequently. N=773.}
    \parbox{0.34\linewidth}{
        \caption{How frequently do you use GenAI to get help with conceptual questions? \label{tab:combined:use:conceptual}}
        \begin{tabular}{lllll}
              & Before       & \%               & After        & \%               \\
        1     & 79           & 10.2\%           & 42           & 5.4\%            \\
        2     & 227          & 29.4\%           & 225          & 29.1\%           \\
        3     & 264          & 34.2\%           & 294          & 38.0\%           \\
        4     & 203          & 26.3\%           & 212          & 27.4\%           
        \end{tabular}
    }
    \hspace{0.03\linewidth}
    \parbox{0.34\linewidth}{
        \caption{How frequently do you use GenAI to get help with debugging? \label{tab:combined:use:debugging}}
        \begin{tabular}{lllll}
              & Before       & \%               & After        & \%               \\
        1     & 144          & 18.6\%           & 111          & 14.4\%           \\
        2     & 281          & 36.4\%           & 291          & 37.6\%           \\
        3     & 211          & 27.3\%           & 244          & 31.6\%           \\
        4     & 137          & 17.7\%           & 127          & 16.4\%           
        \end{tabular}
    }
    \parbox{0.34\linewidth}{
        \caption{How frequently do you use GenAI to get help with homework problems? \label{tab:combined:use:homework}}
        \begin{tabular}{lllll}
              & Before       & \%               & After        & \%               \\
        1     & 182          & 23.5\%           & 179          & 23.2\%           \\
        2     & 340          & 44.0\%           & 341          & 44.1\%           \\
        3     & 187          & 24.2\%           & 187          & 24.2\%           \\
        4     & 64           & 8.3\%            & 66           & 8.5\%            
        \end{tabular}
    }
    \hspace{0.03\linewidth}
    \parbox{0.34\linewidth}{
        \caption{How frequently do you use GenAI to get help with programming projects? \label{tab:combined:use:projects}}
        \begin{tabular}{lllll}
              & Before       & \%               & After        & \%               \\
        1     & 224          & 29.0\%           & 214          & 27.7\%           \\
        2     & 306          & 39.6\%           & 314          & 40.6\%           \\
        3     & 160          & 20.7\%           & 175          & 22.6\%           \\
        4     & 83           & 10.7\%           & 70           & 9.1 \%            
        \end{tabular}
    }
\end{table*}

\subsubsection{Effect Size}

For items with statistically significant pre/post differences, we report a post-hoc effect size using rank-biserial correlation ($r_{rb}$), which complements the Wilcoxon signed-rank test by summarizing the extent to which post responses tend to be higher or lower than pre responses in paired data. Specifically, $r_{rb}$ reflects the balance of positive versus negative paired differences, excluding ties, providing an interpretable measure of both direction and magnitude of the observed shift. For each statistically significant item, the corresponding $r_{rb}$ values fall in the ``large'' range using common benchmarks, indicating that among respondents who changed their rating, the direction of change was predominantly consistent rather than evenly split between increases and decreases. We therefore interpret these effect sizes as evidence of sizable shifts in the distribution of self-reported responses on those items, alongside the statistical significance tests.

\begin{table*}%
    \resizebox{\linewidth}{!}{
    \begin{tabular}{l|ll|ll}
         & \multicolumn{2}{c|}{Wilcoxon Signed Rank} & \multicolumn{2}{c}{Rank Biserial Correlation}\\
        Question                                                               & W & p-value & $r_{rb}$ & Interpretation \\ \hline

        How open are you to using GenAI to get help with homework problems?    & 29882.5  & 0.0140  & 0.9006  & Large  \\
        How open are you to using GenAI to get help with debugging?            & 32020    & 0.6592  &         &        \\
        How open are you to using GenAI to get help with conceptual questions? & 19346    & 0.0008  & 0.9357  & Large  \\
        How comfortable are you using GenAI to help with conceptual questions? & 25323    & <0.0001  & 0.9158  & Large  \\
        How comfortable are you using GenAI to help with homework problems?    & 35840.5  & 0.0118  & 0.8808  & Large  \\
        How comfortable are you using GenAI to help with debugging?            & 35282.5  & 0.0009  & 0.8827  & Large  \\
        How frequently do you use GenAI to get help with programming projects? & 24834.5  & 0.9730  &         &        \\
        How frequently do you use GenAI to get help with debugging?            & 19618    & 0.0239  & 0.9339  & Large  \\
        How frequently do you use GenAI to get help with homework problems?    & 21427    & 0.6806  &         &
    \end{tabular}
    }
    \caption{P-values, Rank Biserial Correlation, and interpretations of significance of change between pre- and post- perception and usage surveys for combined S24 and F24 semesters across CP, DSA-CS, and DSA-DSAI.}
    \label{tab:pvalues_combined_no_ENGR}
\end{table*}

\subsection{Findings from Focus Groups Discussion}\label{sec:results:qual_analysis}

The following findings, drawn from focus groups conducted in Spring and Fall 2024, are organized by research question and build upon the themes described in \Cref{sec:methods:qual_analysis}. \Cref{tab:qualthemes} presents representative excerpts that illustrate each theme.
These themes were developed through an iterative coding process grounded in the student transcripts and are illustrated with representative quotes. We sought to highlight shared patterns, while also recognizing diverse perspectives and occasional disagreement among participants. See Appendix \ref{app:focus_group_questionnaire_and_lab_artifacts} for the focus group protocol.

\subsubsection{RQ1: How and how much are students using GenAI before and after the \ailab intervention?}

\paragraph{From Casual to Strategic Use}
Several students described their early use of GenAI as experimental or based on trial and error, with little structure for prompting. Before the intervention, their use focused on tasks such as summarization, translation, and research. Some expressed frustration with vague prompts or inaccurate responses (code: \textit{frustration with tool limitations} under the theme \textit{From Casual to Strategic Use}):

\begin{quote}
``Usually whenever I try to use ChatGPT, I sort of just give up on it really fast if it doesn't give me the right answer.'' - Focus Group 5, Speaker 3
\end{quote}

Others shared that they had not received any guidance on how to use GenAI productively:

\begin{quote}
``Before we hadn't really been given guidelines on how to use it. . . I thought it was helpful in guiding us, giving different examples of how it could be applied.'' - Focus Group 1, Speaker 3
\end{quote}

After the \ailab, students reported a noticeable shift in their prompting behavior. They began incorporating context, refining prompts, and engaging in more intentional interactions with the tools (code: \textit{prompting strategies}):

\begin{quote}
``Now I give it, for example, everything I do know. . . don't change those answers. Based on what I have, give me the answer to A.'' - Focus Group 6, Speaker 2
\end{quote}

\begin{quote}
``I felt like it did much better once it was given clear rules. . . especially if I would explicitly dictate what I wanted it to tell me.'' - Focus Group 5, Speaker 1
\end{quote}

These comments reflect a broader theme of moving from casual or inefficient usage toward more strategic application of GenAI, even if the frequency of use did not always change.

\paragraph{Stable Frequency, Evolving Rationale.} Although prompting behavior improved, many participants reported that the \textit{amount} of usage remained relatively stable. Still, they described using the tools more effectively and with clearer goals (code: \textit{usage rationale}):

\begin{quote}
``It's pretty much the same. I just learned better ways to use it.'' - Focus Group 1, Speaker 3
\end{quote}

\begin{quote}
``I don't think I use it more, but I definitely use it better. Now I know how to refine my prompts instead of wasting time rewriting them over and over.'' - Focus Group 3, Speaker 2
\end{quote}

A few students reported increased usage after the intervention, connecting this with their improved skillset:

\begin{quote}
``I had used it sparsely... after that I found myself using it maybe twice a week for a variety of different classes.'' - Focus Group 5, Speaker 1
\end{quote}

In contrast, some students consciously reduced their use, citing concerns about overreliance or a desire to preserve their own learning process (code: \textit{avoidance and dependency concerns}):

\begin{quote}
``At some point, I felt like I was relying too much on ChatGPT. I didn't want it to replace my own thinking, so I started using it less.'' - Focus Group 4, Speaker 3
\end{quote}

These reflections suggest that even when frequency remained unchanged, the rationale behind GenAI use became more intentional and informed.

\paragraph{Improved Understanding of Limitations.} Many students became more skeptical and critical of GenAI's limitations, especially in areas like math, statistics, or coding. These reflections align with the theme \textit{Improved Understanding of Limitations} and codes related to \textit{skepticism and error awareness}:

\begin{quote}
``A lot of the time in certain areas like statistics or linear algebra, it gives incorrect answers. It's obviously incorrect.'' - Focus Group 1, Speaker 3
\end{quote}

\begin{quote}
``It would make it look really, really right if you lack the sense of judgment. . . I couldn't tell whether you're accurate or not.'' - Focus Group 5, Speaker 4
\end{quote}

\begin{quote}
``[GenAI] just told me... [but] I cannot access articles through links. It was just funny how... really obvious things beyond its capability.'' - Focus Group 6, Speaker 2
\end{quote}

Together, these comments highlight a growing awareness of GenAI's boundaries, signaling a shift from passive use to more critical engagement.

Overall, while students' frequency of GenAI use varied, the \ailab intervention appeared to influence how they approached and evaluated these tools. Across courses, participants became more intentional in how they prompted, more critical of AI responses, and more reflective about when and whether to rely on GenAI in their learning.

\subsubsection{RQ2: Are Students Open to Using GenAI in Their Learning? If So, In What Scenarios?}

To explore this question, we examined how students described their willingness to use GenAI and the types of academic situations where they found it helpful or avoided it. Their responses revealed a general openness to the use of tools, thoughtful boundaries, and internal tensions.

\paragraph{AI as a Support Tool for Learning}
Many students described using GenAI as a way to scaffold understanding, clarify unfamiliar concepts, or explore new ideas. Rather than treating it as a shortcut, several participants emphasized that they used GenAI as a starting point, which is something closer to a study partner or TA rather than a solution generator:

\begin{quote}
 ``I basically use it to break down hard concepts into simpler terms.'' - Focus Group 6, Speaker 3
 \end{quote}
\begin{quote}
 ``It's sort of like questions I'd ask a TA. I just ask Chat first.'' - Focus Group 3, Speaker 5
 \end{quote}
\begin{quote}
 ``I primarily use it for research... it's been really good for taking notes.'' - Focus Group 5, Speaker 2
 \end{quote}
\begin{quote}
 ``It becomes useful in your programming... I'll ask, what's an example code using this documentation?'' - Focus Group 3, Speaker 4
 \end{quote}

These comments suggest that for many students, GenAI served as an academic aid, helping them build confidence, test ideas, or organize information while still keeping them actively involved in the learning process.

\paragraph{Setting Boundaries: When Students Chose Not to Use GenAI}

Despite the benefits, students were also quick to name situations where they intentionally avoided using GenAI. Some mentioned courses or topics, particularly math-heavy assignments or personal writing, where they felt the tool lacked accuracy or contextual understanding. Others expressed concerns about academic integrity or detection by instructors:

\begin{quote}
 ``It's not super great at math, so any kind of probability stuff, linear algebra, calculus... not too great at.'' - Focus Group 1, Speaker 2
 \end{quote}
\begin{quote}%
 ``If your code is very complex, sometimes it just doesn't know where to start.'' - Focus Group 5, Speaker 2
 \end{quote}
\begin{quote}
 ``Probably on projects or homework... there's a risk of an AI detection being run.'' - Focus Group 1, Speaker 3
 \end{quote}
\begin{quote}
 ``It just sounds awkward and robotic.'' - Focus Group 5, Speaker 3
 \end{quote}

These examples highlight that students were not blindly adopting GenAI; instead, they exercised discernment by choosing to use it selectively and in ways that felt appropriate given the context, stakes, and their own values.

\paragraph{Ethical Concerns and Internal Tension}

A recurring theme in students' reflections was the tension between leveraging GenAI to support their thinking and maintaining a sense of personal ownership over their learning. While some appreciated the clarity and productivity boost that GenAI offered, others worried that it might compromise their growth or even alter their perception of themselves as learners:

\begin{quote}
    ``I don't want it to just generate content for me. I want to make sure I'm still learning.'' - Focus Group 1, Speaker 3
    
    \noindent``Some professors let us use it if we document everything. Others just say no AI at all.'' - Focus Group 5, Speaker 2
    
    \noindent``I reduced using it... I just feel like I'm getting more dumb... a puppet of ChatGPT.'' - Focus Group 5, Speaker 4
\end{quote}

These tensions point to a deeper level of reflection among students, not only about the effectiveness of GenAI but also about what constitutes authentic learning, fairness, and academic identity. While views varied, it was clear that many were still negotiating how to integrate GenAI into their academic lives in a responsible manner.

\paragraph{Closing Reflection}
Across the themes above, focus group participants described being open to using GenAI while also articulating norms of restraint, verification, and boundary-setting. Participants reported experimenting with GenAI, learning to provide more context and iterate on prompts, and becoming more skeptical of correctness -- often describing cross-checking outputs against course materials, tests, or their own reasoning. Many framed GenAI as a support tool for understanding and debugging rather than a substitute for completing graded work, and several voiced an ongoing tension between convenience and maintaining ownership of their learning. These qualitative accounts are intended to contextualize the survey patterns and to illustrate how students described their decision-making about GenAI use; broader study limitations and threats to validity are discussed in \Cref{sec:discussion:threats}.

\subsection{Additional Descriptive Results}

\begin{figure*}%
    \centering
    \begin{subfigure}{0.49\linewidth}
        \centering
        \includegraphics[width=\linewidth]{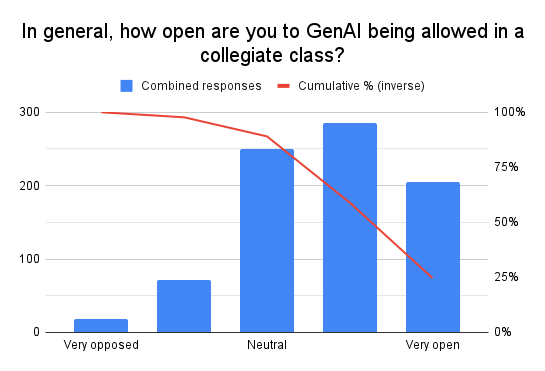}
        \caption{Openness to GenAI being allowed in a collegiate class, for CS and ENGR students, from the post-perception survey. There was no significant difference between pre- and post-perception surveys on this question.}
        \label{fig:open_to_genai_being_allowed}
    \end{subfigure}
    \hfill
    \begin{subfigure}{0.49\linewidth}
        \centering
        \includegraphics[width=\linewidth]{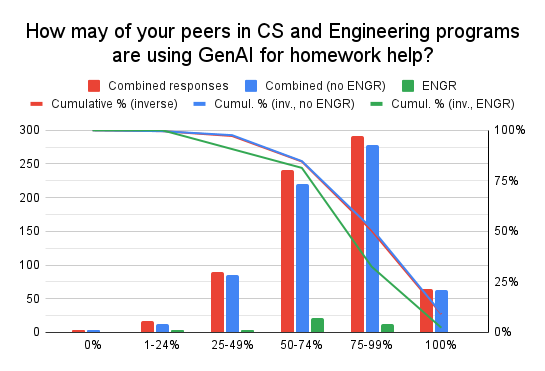}
        \caption{``No Idea'' responses excluded.  They compose the following amounts: Combined: 121/826 (14.65\%), CS only: 111/773 (14.36\%), ENGR only: 10/53 (18.87\%). Dropping ENGR from the combined numbers changes any one percentage by no more than 0.95\%.}
        \label{fig:how_many_peers}
    \end{subfigure}
    \caption{(a) Post-perception survey results for openness to GenAI being allowed in a collegiate class and (b) Pre-usage survey results for perceived peer usage of GenAI for homework help.}
    \Description{Left: Openness to GenAI being allowed in a collegiate class, for CS and ENGR students, with inverse cumulative distributions. Right: ``No Idea'' responses excluded.  They compose the following amounts: Combined: 121/826 (14.65\%), CS only: 111/773 (14.36\%), ENGR only: 10/53 (18.87\%). Dropping ENGR from the combined numbers changes any one percentage by no more than 0.95\%.}
\end{figure*}

In this subsection, we present several supportive (non-paired) observations that illuminate broader student behavior and attitudes.
The first of these was the perception survey question, \textit{``In general, how open are you to GenAI being allowed in a collegiate class?''}  As seen in \Cref{fig:open_to_genai_being_allowed}, 89.05\% of students are neutral or open to GenAI being allowed in a collegiate class, with 58.97\% being open to it.  As discussed in \Cref{sec:results:qual_analysis}, focus groups showed a highly nuanced and wide array of perspectives from undergraduate students as a whole, including \textit{all} of the perspectives we have heard from faculty. %

The second non-paired question we present is from the pre-usage survey, \textit{``How many of your peers in CS and Engineering programs are using GenAI for homework help?''}.  Interestingly, as seen in \Cref{fig:how_many_peers}, CS sophomores (on average) and Engineering freshmen from nearly completely disjoint populations report nearly the same distribution of perceived usage of GenAI on homework by their peers.
Furthermore, comparing perceived usage to actual usage \textit{before} the intervention (as the peer question was before the intervention), as in \Cref{tab:combined:use:homework}, only around 50\% of students predicted correctly.  Specifically, S24 and F24, 31.60\% of the students reported moderate or very frequent use on homework, with 75.42\% reporting using it on occasion or more, but 84.54\% reported $\ge 50\%$ of their peers used it for homework and only 50.36\% reported $\ge 75\%$ of their peers were using it for homework, which means that only half of the students had it right, indicating that students perceive less of their peers using it than actually do.

\begin{figure*}
    \centering
    \caption{Self-reported change in desire to use GenAI}
    \begin{subfigure}[b]{0.49\linewidth}
        \centering
        \caption{Combined across all courses and semesters.}
        \includegraphics[width=\linewidth]{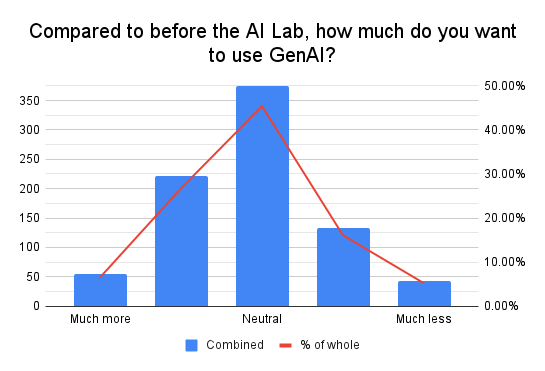}
        \label{fig:compared_to_before_only_combined}
        \Description{Self-reported change in desire to use GenAI, combined across all courses and semesters.  Skewed normal, towards more usage post.}
    \end{subfigure}
    \hfill
    \begin{subfigure}[b]{0.49\linewidth}
        \centering
        \caption{Normalized, shown for each individual course.}
        \includegraphics[width=\linewidth]{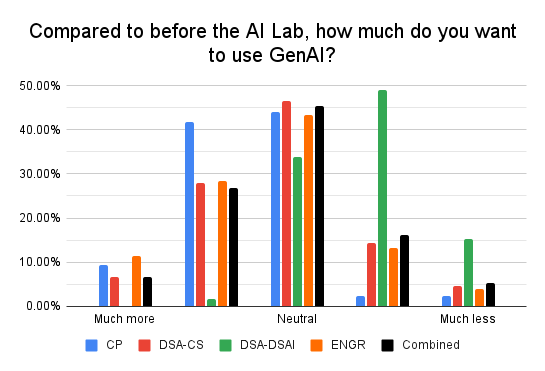}
        \label{fig:compared_to_before_all_combined}
        \Description{Self-reported change in desire to use GenAI, combined across all courses and semesters.  Skewed normal, for most courses.}
    \end{subfigure}
\end{figure*}

To round out our non-paired question analysis, we asked the students to self-evaluate how much their desire to use GenAI changed because of the intervention.  The combined results across all four courses are shown in \Cref{fig:compared_to_before_only_combined} and \Cref{fig:compared_to_before_all_combined}.  The authors note that removing ENGR changes none of the \Cref{fig:compared_to_before_only_combined} percentages more than 0.33\% (thus the conclusions drawn from this chart are applicable either way).  Further discussion can be found in \Cref{sec:discussion:diff_in_courses}

\section{Discussion}\label{sec:discussion}
This study evaluated \emph{\ailab}, a brief, course-embedded GenAI intervention deployed across multiple early undergraduate courses over two semesters. Our aim was not to determine whether GenAI improves learning outcomes or to measure tool usage directly, but to examine whether participation in a structured, scaffolded lab was associated with shifts in students' self-reported GenAI practices and scenario-specific attitudes (openness and comfort), and to use post-intervention focus groups to contextualize how students described engaging with GenAI.

Across the paired surveys in the CS courses, we observed a consistent pattern: students reported higher comfort using GenAI across several common academic scenarios, and higher openness in some scenarios, while reported frequency of GenAI use on graded work (homework and programming projects) did not increase. The primary reported usage change was an increase in GenAI use for debugging. This combination is important because it suggests that, in this setting, the intervention's apparent influence was more visible in students' reported readiness and confidence to engage with GenAI -- and in where they reported using it -- than in a broad increase in reported use on assessed work.
Focus group accounts add depth to this pattern by clarifying what ``change'' looked like from students' perspectives. Participants described shifting from one-shot or minimally specified prompts toward more strategic interactions (e.g., providing more context, iterating on prompts, and explicitly constraining what they wanted the tool to do). They also described becoming more skeptical of correctness and more likely to verify outputs (e.g., by cross-checking against course materials, tracing logic, or testing code). Importantly, participants articulated clearer boundaries for when GenAI felt appropriate -- often framing it as a TA-like support for understanding concepts or debugging -- while expressing hesitation about relying on it to produce complete solutions in higher-stakes, graded contexts. Some participants also voiced internal tension about dependence, describing a desire to benefit from GenAI's efficiency without ``outsourcing'' their own problem solving. Taken together, these accounts suggest that the \ailab may shape the strategies and decision-making students bring to GenAI use, even when self-reported frequency on graded work remains stable.

Interpretation is bounded by the threats to validity in \Cref{sec:discussion:threats}, particularly reliance on self-report and the absence of behavioral or learning-outcome measures.
The remainder of this Discussion examines how observed patterns vary across courses and semesters, interprets the null findings that bound what changed, and revisits common instructor concerns about whether GenAI instruction increases inappropriate use in graded work within the limits of our measures.

\subsection{Differences in Courses} \label{sec:discussion:diff_in_courses}
Although our primary analyses focus on the three CS courses, we also deployed the \ailab in a first-year engineering course (ENGR). In ENGR, a larger number of survey items showed detectable pre/post shifts than in the CS offerings (\Cref{tab:number_of_significant_pvalues}). We treat this course-by-course comparison as exploratory, since the number of statistically significant items can be influenced by baseline response distributions and sample size. Nevertheless, the pattern suggests that students' responses to a short, scaffolded GenAI intervention may vary by disciplinary and curricular context.

To examine whether ENGR meaningfully altered the pooled interpretation for Fall 2024, we repeated the paired analyses with and without ENGR included (\Cref{tab:pvalues_f24_comparison}). The overall pattern of results was stable, but one inference changed: comfort using GenAI for homework reached significance only when ENGR was included in the Fall 2024 pool. This shift is consistent with ENGR students showing stronger pre/post movement on several items, and it highlights that course context may function as a moderator of intervention effects rather than assuming a single ``average'' impact across offerings. Sample sizes for this data can be found in \Cref{tab:combined_perception_semester_counts,tab:combined_usage_semester_counts}.

Finally, students' self-reported change in desire to use GenAI after the \ailab differed by course (\Cref{fig:compared_to_before_all_combined}). CP, DSA-CS, and ENGR showed distributions tilted toward increased desire, whereas DSA-DSAI showed a comparatively larger share reporting decreased desire. We do not interpret these differences as evidence of better or worse outcomes across courses; instead, they motivate future work to test which contextual factors (e.g., prior exposure to GenAI, perceived stakes, course policies, and the structure of the post-lab task) help explain why the same scaffold produces different reported shifts across settings.

\begin{table*}%
    \begin{tabular}{lllll}
                                       & CP & DSA-CS & DSA-DSAI & ENGR\\
        Number of significant p-values & 2  & 4      & 4        & 6
    \end{tabular}
    \caption{Exploratory summary of within-course pre/post shifts: number of survey items (out of 9) with p<.05. This count is sensitive to baseline distributions and sample size and should be interpreted as descriptive comparison rather than a measure of effect magnitude.\label{tab:number_of_significant_pvalues}}
\end{table*}

\begin{table*}%
    \resizebox{\linewidth}{!}{
        \begin{tabular}{l|ll|ll|ll|ll}
            &  \multicolumn{4}{c|}{F24 With ENGR}        & \multicolumn{4}{c}{F24 Without ENGR} \\
            &  \multicolumn{2}{c|}{Wilcoxon Signed Rank} & \multicolumn{2}{c|}{Rank Biserial Corr.} & \multicolumn{2}{c|}{WSR} & \multicolumn{2}{c}{RBC}\\
            Question                                    & W & p-value & $r_{rb}$ & Interpretation & W  & p-value & $r_{rb}$ & Interp.\\ \hline
            
            Open to using for homework problems?        & 20420   & 0.0011  & 0.9014 & Large   & 16905.5 & 0.0047  & 0.9030 & Large \\
            Open to using for debugging?                & 15433.5 & 0.0001  & 0.9255 & Large   & 14131   & 0.0012  & 0.9189 & Large \\
            Open to using for conceptual questions?     & 13353.5 & <0.0001 & 0.9355 & Large   & 10819   & 0.0002  & 0.9379 & Large \\
            Comfortable using for conceptual questions? & 17965   & <0.0001 & 0.9132 & Large   & 15549   & 0.0002  & 0.9114 & Large \\
            Comfortable using for homework problems?    & 24981.5 & 0.0098  & 0.8793 & Large   & 21701   & 0.0592  &        &       \\
            Comfortable using for debugging?            & 20500   & <0.0001 & 0.9010 & Large   & 18270.5 & <0.0001 & 0.8952 & Large \\
            Frequency of use on programming projects?   & 20098   & 0.6582  &        &         & 17090.5 & 0.9096  &        &       \\
            Frequency of use on debugging?              & 14054   & 0.0045  & 0.9315 & Large   & 12310.5 & 0.0370  & 0.9287 & Large \\
            Frequency of use on homework problems?      & 15276   & 0.7413  &        &         & 13443   & 0.4879  &        &       
        \end{tabular}
    }
    \caption{Fall 2024 Wilcoxon signed-rank and rank-biserial results for CP and DSA-CS, shown with and without ENGR.\label{tab:pvalues_f24_comparison}}
\end{table*}

\begin{table*}
    \parbox{0.5\linewidth}{
        \centering\captionsetup{justification=centering}
        \begin{tabular}{llll}
        \textbf{Perception}  & \textbf{S24}  & \textbf{F24}  & \textbf{Grand Total} \\
        CP                   & --            & 43            & 43                   \\
        DSA-CS               & 129           & 547           & 676                  \\
        DSA-DSAI             & 59            & --            & 59                   \\\hline
        \textbf{Total}       & \textbf{188}  & \textbf{590}  & \textbf{778}         \\
        ENGR                 & --            & 53            & 53                   \\\hline
        \textbf{Grand Total} & \textbf{188}  & \textbf{643}  & \textbf{831}
        \end{tabular}
        \caption{Perception survey totals by course and semester. \label{tab:combined_perception_semester_counts}}
    }
    \parbox{0.49\linewidth}{
        \centering\captionsetup{justification=centering}
        \begin{tabular}{llll}
        \textbf{Usage}       & \textbf{S24}  & \textbf{F24}  & \textbf{Grand Total} \\
        CP                   & --            & 43            & 43                   \\
        DSA-CS               & 127           & 544           & 671                  \\
        DSA-DSAI             & 59            & --            & 59                   \\\hline
        \textbf{Total}       & \textbf{186}  & \textbf{587}  & \textbf{773}         \\
        ENGR                 & --            & 53            & 53                   \\\hline
        \textbf{Grand Total} & \textbf{186}  & \textbf{640}  & \textbf{826}
        \end{tabular}
        \caption{Usage survey totals by course and semester. \label{tab:combined_usage_semester_counts}}
    }
\end{table*}

\subsection{Differences between semesters}\label{sec:discussion:diff_in_sem}
We conducted an exploratory comparison of Spring 2024 (S24) and Fall 2024 (F24) to examine whether students' self-reported GenAI practices differed across terms. These comparisons should be interpreted cautiously because the semester samples differ in size and course composition (e.g., S24 includes DSA-DSAI while F24 includes CP and ENGR, and DSA-CS represents a larger share of responses in F24). As a result, observed differences may reflect cohort and course-context differences in addition to any broader temporal shifts.

\begin{figure*}
    \centering
    \includegraphics[width=0.5\linewidth]{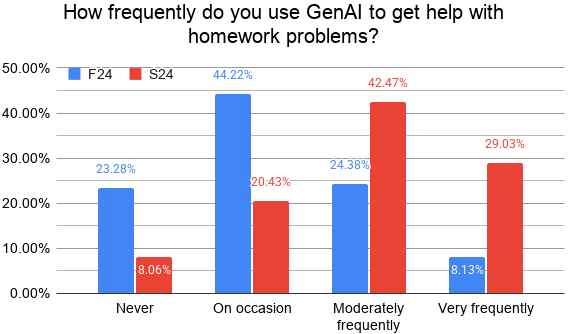}
    \caption{A stark difference between Fall 2024 and Spring 2024, in percentage of students who used GenAI moderately or very frequently compared to never or on occasion.  F24 shows much higher percentages in never or on occasion (67.50\%) compared to S24 (28.49\%).  Note that populations were primarily composed of DSA-CS students in both semesters.  See \Cref{sec:discussion:diff_in_sem} for analysis.}
    \label{fig:frequency_homework_comparingFallAndSpring}
    \Description{A stark difference between Fall 2024 and Spring 2024, in percentage of students who used GenAI moderately or very frequently compared to never or on occasion.  F24 shows much higher percentages in never or on occasion (67.50\%) compared to S24 (28.49\%).  Note that populations were primarily composed of DSA-CS students in both semesters.  See \Cref{sec:discussion:diff_in_sem} for analysis.}
\end{figure*}

The clearest between-semester difference appeared on the usage item asking how frequently students use GenAI to get help with homework problems. In the pre-intervention survey, S24 respondents reported higher homework-use frequency than F24 respondents (\Cref{fig:frequency_homework_comparingFallAndSpring}), with F24 responses more concentrated in the ``Never'' and ``On occasion'' categories. Visually, this between-semester difference appears to persist in the post-intervention distributions as well, suggesting that the \ailabshort did not coincide with a large shift in this specific self-reported behavior at the semester level.

Focus group accounts from both semesters provide complementary context for this pattern. Across S24 and F24, participants discussed integrity- and learning-oriented boundaries (e.g., when GenAI felt appropriate, verification practices, and concerns about dependence), and several F24 participants explicitly referenced course or instructor AI policies when describing their decisions about using GenAI on graded work.%

Because our study was not designed to isolate why baseline self-reports differ across semesters, we do not attribute this pattern to any single cause. Plausible contributing factors include differences in course contexts and instructor policies/messaging, as well as rapidly evolving student familiarity with GenAI tools. Future work tracking the same courses over multiple terms and triangulating self-reports with behavioral traces will be necessary to determine whether these baseline shifts reflect changes in actual tool use, reporting norms, or both.

\subsection{What We Didn't See (Null Findings)}
In this subsection, we focus on the paired survey analyses for the CS courses (DSA-CS, DSA-DSAI, and CP) and highlight outcomes that did not show statistically detectable pre/post shifts after the \ailab.

\paragraph{Null findings as informative outcomes.}
Three outcomes did not show pre/post change in the CS-only paired analyses: students' self-reported frequency of using GenAI for homework and for programming projects, and their self-reported openness to using GenAI for debugging (\Cref{tab:combined:use:homework,tab:combined:use:projects,tab:combined:open:debugging}). These null results are pedagogically consequential: instructors often worry that acknowledging GenAI in class may increase reliance on GenAI for graded work, so stable self-reported frequency on homework and projects helps bound what changed and what did not in students' reported practices. At the same time, because these measures are self-reported and we did not independently verify tool use, we interpret stability as evidence about students' reported behavior rather than as evidence about actual behavior.

\paragraph{Why debugging openness did not change.}
Reported openness to using GenAI for debugging started high before the intervention (\Cref{tab:combined:open:debugging}), leaving limited room for further increases. In addition, the \ailab was not primarily designed to encourage or discourage students using GenAI for debugging; instead, it was designed to help students experience benefits and limitations of GenAI and to practice verification and boundary-setting in course-relevant contexts. This framing aligns with the fact that openness increased for conceptual questions and homework scenarios, where the intervention more directly targeted students' decision-making (\Cref{tab:combined:open:conceptual,tab:combined:open:homework}). One plausible interpretation is that students already viewed debugging as an appropriate context for GenAI assistance (e.g., lower perceived stakes or perceived integrity risk than graded submissions), but our data do not allow us to determine the cause of the high baseline openness. Notably, while openness to debugging did not change, students' reported comfort using GenAI for debugging and their reported frequency of using GenAI for debugging did increase (\Cref{tab:combined:comfort:debugging,tab:combined:use:debugging}), suggesting that students may have shifted how they engage with GenAI while debugging even if their willingness was already high.

\paragraph{Why homework and project frequency did not change.}
We also did not observe statistically significant changes in students' self-reported frequency of GenAI use for homework problems or programming projects (\Cref{tab:combined:use:homework,tab:combined:use:projects}). Given the large paired sample, we would expect to detect a broad, consistent shift in reported GenAI use on homework/projects if one occurred. We did not observe such a shift, suggesting the intervention was not associated with a widespread increase in graded-work use.
This pattern is consistent with the possibility that course policies, assessment stakes, or students' existing norms around graded work constrain changes in reported frequency over a short intervention. However, our design does not allow us to distinguish among these explanations. Focus groups contained accounts ranging from restraint to selective use in graded contexts, but because participation was self-selected and not intended to estimate prevalence, we do not treat these accounts as evidence about how common any particular practice was.

\paragraph{Summary.}
Overall, these null findings help bound the observed effects of the \ailab: the intervention was associated with shifts in reported comfort and in some scenario-specific attitudes and practices, but it was not associated with broad increases in self-reported GenAI use on homework or programming projects in the paired CS sample.

\subsection{If I teach them GenAI, will they use it to cheat more?}
A common concern is that explicit GenAI instruction may increase inappropriate reliance on GenAI for graded work. In our paired surveys, students did not report increased frequency of using GenAI for homework or programming projects following the \ailab. This pattern is compatible with the possibility that structured discussion of limitations and boundaries does not increase students' reported reliance in graded contexts. However, we do not treat this as evidence about actual misconduct: our measures are self-reported, and we did not independently verify GenAI usage or evaluate the provenance of submitted work. We therefore frame this finding as an informative signal for instructors -- one that should be corroborated in future work with behavioral evidence.

\subsection{Threats to Validity}\label{sec:discussion:threats}
This study uses a single-group pre/post design, so observed differences should be interpreted as associations with participation in the \ailab rather than causal effects.

Our primary outcomes (frequency, openness, and comfort) are self-reported and may be affected by recall error, ambiguity about what counts as ``using GenAI,'' and social desirability, particularly for items involving graded work and integrity norms. We did not collect behavioral traces of GenAI use (e.g., tool logs or chat histories suitable for analysis), so we cannot verify whether reported frequency matches actual usage. We also did not measure learning outcomes or code quality, so the study cannot determine whether the intervention improved learning or reduced inappropriate use.

Implementation and context varied across offerings (e.g., prompts, timing, post-lab tasks, and instructor policy messaging), which may moderate effects and limit transfer to other institutions or course types. Focus groups were voluntary and conducted weeks after the intervention; group dynamics and retrospective accounts may shape what participants report. We therefore treat qualitative themes as illustrative accounts that contextualize survey patterns rather than prevalence estimates. Finally, because we tested multiple survey items, we emphasize the overall pattern of results and effect sizes and interpret individual p-values cautiously.

\subsection{Conclusions and Future Outlooks}
We evaluated \ailab, a brief, scaffolded GenAI lab embedded in multiple early undergraduate courses, using paired pre/post surveys and post-intervention focus groups. Across the paired CS analyses, students reported increased comfort using GenAI across common academic scenarios and increased openness in specific contexts (notably conceptual questions and homework help), while self-reported frequency of GenAI use on homework and programming projects remained stable. The primary reported usage change was increased GenAI use for debugging. Focus group themes provide complementary context: participants described adopting more iterative and strategic prompting, becoming more skeptical of correctness, and articulating clearer boundaries around integrity and dependence. Taken together, these findings suggest that short, adoptable scaffolds can shape students' reported readiness and strategies for engaging with GenAI without a broad increase in self-reported GenAI use on graded work.

Looking ahead, several directions would strengthen both the empirical basis and the practical guidance for instructors. First, future work should triangulate self-report with behavioral evidence (e.g., usage traces, submitted interaction artifacts where feasible) and with learning measures (e.g., code quality, conceptual understanding, or performance on assessments designed to evaluate independent problem solving). Second, longitudinal deployments that track the same courses across additional terms would help disentangle intervention effects from shifting baseline norms and policies as GenAI becomes more integrated into students' academic lives. Finally, replications with larger and more diverse student populations and in additional disciplines can clarify which elements of the \ailab structure transfer readily and which require adaptation to local course constraints, assessment practices, and integrity expectations.

\section{Appendix}
\subsection{Focus Group Questionnaire and \ailab Artifacts}\label{app:focus_group_questionnaire_and_lab_artifacts}
Due to space constraints, the focus group questionnaire and recruitment protocol are available anonymously \href{https://docs.google.com/document/u/0/d/e/2PACX-1vTcMp63uKJnlOsoXOKEGbd9K7ZHlaD-tujtUUNqcZumX9hBXqbBvT2ekdl67B85YjDc91VbVaDpxK6F/pub}{(\underline{click here})}. Anonymized versions of a pre-lab, in-class outline, post-lab, and other instructional materials are also available below the focus group information at the same link.

\section*{Acknowledgments}
This work was funded by Purdue's Innovation Hub (IH-AI-23002). The authors acknowledge the Teaching Assistants from courses CS 211, CS 251, and CS 253 during Spring and Fall of 2024, who graded the \ailab assignment portion and managed questions. The authors would also like to thank Sean Brophy for being the first to adopt the \ailab outside of CS and for being gracious enough to allow them to collect data.  The authors finally thank Purdue's Institutional Data Analytics and Assessment (IDA+A) as well as the Center for Instructional Excellence (CIE), in particular Brooke Harris-Thomas and David Nelson, both of whom were instrumental in the success of this project.

This study was conducted in accordance with protocol IRB-2023-1079.
\bibliographystyle{ACM-Reference-Format}
\bibliography{refs}

\end{document}